\documentclass[%
 longbibliography,aip,
 amsmath,amssymb,
reprint,%
]{revtex4-1}

\usepackage{graphicx}
\usepackage{dcolumn}
\usepackage{bm}
\usepackage{amsmath}
\usepackage[utf8]{inputenc}
\usepackage[T1]{fontenc}
\usepackage{mathptmx}
\usepackage{upgreek}
\usepackage{xcolor}

\usepackage{tabularx}
\usepackage{booktabs}

\usepackage{dcolumn}
\usepackage{siunitx}
\DeclareSIUnit\year{a}
\DeclareSIUnit\ppm{ppm}
\usepackage{tikz}
\usetikzlibrary{decorations.pathmorphing,shapes}

\makeatletter
\def\@bibdataout@aps{%
\immediate\write\@bibdataout{%
@CONTROL{%
apsrev41Control%
\longbibliography@sw{%
    ,author="08",editor="1",pages="0",title="0",year="1"%
    }{%
    ,author="08",editor="1",pages="0",title="",year="1"%
    }%
  }%
}%
\if@filesw \immediate \write \@auxout {\string \citation {apsrev41Control}}\fi 
}
\makeatother

\newcounter{sarrow}

\begin{document}

\preprint{}

\title[High Voltage Determination and Stabilization]{High Voltage Determination and Stabilization for Collinear Laser Spectroscopy Applications}

\author{Kristian~K\"onig}
  \email{kkoenig@ikp.tu-darmstadt.de}
  \affiliation{Institut für Kernphysik, Department of Physics, Technische Universität Darmstadt, 64289 Darmstadt, Germany}
  \affiliation{Facility for Rare Isotope Beams, Michigan State University, East Lansing, MI 48824, USA}
\author{Finn K\"ohler}%
  \affiliation{Institut für Kernphysik, Department of Physics, Technische Universität Darmstadt, 64289 Darmstadt, Germany}
\author{Julian Palmes}%
  \affiliation{Institut für Kernphysik, Department of Physics, Technische Universität Darmstadt, 64289 Darmstadt, Germany}
\author{Henrik Badura}%
  \affiliation{Physikalisch-Technische Bundesanstalt, 38116 Braunschweig, Germany}  
\author{Adam Dockery}
  \affiliation{Facility for Rare Isotope Beams, Michigan State University, East Lansing, MI 48824, USA}
  \affiliation{Department of Physics and Astronomy, Michigan State University, East Lansing, MI 48824, USA}
\author{Kei Minamisono}
  \affiliation{Facility for Rare Isotope Beams, Michigan State University, East Lansing, MI 48824, USA}
  \affiliation{Department of Physics and Astronomy, Michigan State University, East Lansing, MI 48824, USA}
\author{Johann Meisner}%
  \affiliation{Physikalisch-Technische Bundesanstalt, 38116 Braunschweig, Germany}
\author{Patrick~Müller}
  \affiliation{Institut für Kernphysik, Department of Physics, Technische Universität Darmstadt, 64289 Darmstadt, Germany}
\author{Wilfried Nörtershäuser}
  \affiliation{Institut für Kernphysik, Department of Physics, Technische Universität Darmstadt, 64289 Darmstadt, Germany}
  \affiliation{Helmholtz Forschungsakademie Hessen für FAIR (HFHF), Campus Darmstadt, Schlossgartenstr.~9, 64289 Darmstadt, Germany}
\author{Stephan Passon}%
  \affiliation{Physikalisch-Technische Bundesanstalt, 38116 Braunschweig, Germany}

\date{\today}

\begin{abstract}
Fast beam collinear laser spectroscopy is the established method to investigate nuclear ground state properties such as the spin, the electromagnetic moments, and the charge radius of exotic nuclei. These are extracted with high precision from atomic observables, i.e., the hyperfine splitting and its the isotope shift, which becomes possible due to a large reduction of the Doppler broadening by compressing the velocity width of the ion beam through electrostatic acceleration. With the advancement of the experimental methods and applied devices, e.g., to measure and stabilize the laser frequency, the acceleration potential became the dominant systematic uncertainty contribution. To overcome this, we present a custom-built high-voltage divider, which was developed and tested at the German metrology institute (PTB), and a feedback loop that enabled collinear laser spectroscopy to be performed at a 100-kHz level. Furthermore, we describe the impact of field penetration into the laser-ion-interaction region. This strongly affects the determined isotope shifts and hyperfine splittings, if Doppler tuning is applied, i.e., the ion beam energy is altered instead of scanning the laser frequency. 
Using different laser frequencies that were referenced to a frequency comb, the field penetration was extracted laser spectroscopically. This allowed us to define an effective scanning potential to still apply the faster and easier Doppler tuning without introducing systematic deviations.
\end{abstract}

\maketitle

\section{Introduction}
Long-term stable high-voltage sources are key for a variety of modern high-precision experiments in particle physics, e.g., for the determination of the neutrino mass by measuring the electron-endpoint energy \cite{Katrin.2022, Arenz.2018} or for the operation of large-scale calorimeters \cite{Bartoloni.2007}, nuclear physics, e.g., the determination of masses \cite{Boehm.2016, Cavenago.2021} or charge radii \cite{Krieger.2011}, atomic physics, e.g., in storage ring experiments \cite{Ullmann.2017, Yan.2023}, as well as in the search for physics beyond the standard model by measuring the electron electric dipole moment \cite{Bishof.2016}. Although those fields and their experimental approaches are diverse, they depend on stable high voltages to accelerate charged particles or to define trap potentials.
In this work, we focus on the application to collinear laser spectroscopy, which probes electrons in the atomic shell. This technique enables the measurement of exotic systems like highly-charged particles to test the QED in extreme fields \cite{Ullmann.2017, Lochmann.2014} or to investigate nuclear ground state properties of short-lived isotopes \cite{Campbell.2016,Yang23}. Thanks to the advances in laser frequency measurements, a relative accuracy of $10^{-9}$ can be easily achieved with modern wavelength meters or frequency combs, which is sufficient for these applications \cite{Verlinde.2020,Koenig.2020b}. Nowadays, the systematic limitation of this experimental approach is the measurement of the beam energy, and hence, the precise determination of the electrostatic acceleration potential. Since field penetrations, space charge effects and contact voltages, which can hardly be quantified, add to the applied acceleration potential, the kinetic beam energy can be determined most precisely by calibration measurements with reference isotopes \cite{Koenig.2021}. However, long-term measurements strongly rely on stable conditions, and are mostly limited by drifting high-voltages. In this work, we present our approach to measure these voltages and apply a stabilization scheme to reduce drifts to a 1-ppm level. Similar approaches, which have been developed for different high-precision physics experiments, are described in Ref. \cite{Chen.2008, Thuemmler.2009,Schury.2020, Fischer.2021, Miwa.2023}. Furthermore, we present a procedure to quantify field penetration into the laser-ion interaction region since this critically affects the commonly applied voltage scans to match the resonance condition between Doppler-shifted laser frequency and atomic transition.

\section{Setup}
In collinear laser spectroscopy experiments an ion beam with a kinetic energy of a few 10\,keV is used, which has been electrostatically accelerated by floating the ion source on a high-voltage potential. The ion beam is then superposed with a laser beam and by scanning the laser frequency in the rest-frame of the ion, an electronic transition is probed. At resonance, the ions are excited and emit fluorescence light that can be detected with photo-multiplier tubes. 
Due to the Doppler effect, the resonance frequency $\nu_0$ in the rest-frame of the ion differs from the laser frequency $\nu_\mathrm{lab}$ measured in the laboratory frame according to
\begin{equation}
    \nu_0 = \nu_\mathrm{lab} \gamma (1-\beta \cos(\alpha)),
    \label{eq:CLS}
\end{equation}
with the time-dilation factor $\gamma=(1-\beta^2)^{-1/2}$, the velocity $\beta$ in units of the speed of light and the angle $\alpha$ between ion and laser beam, which is $\alpha=0^\circ$ for collinear and $\alpha=180^\circ$ for anticollinear geometry. To extract the rest-frame transition frequency, the kinetic energy and, hence, the acceleration potential, needs to be precisely known. This is challenging due to large uncertainties of the starting potential in the ion source, contact voltages and contributions caused by field penetration into the laser-ion interaction region.
Luckily, in relative measurements for the determination of differential charge radii from isotope shifts or electromagnetic moments from the hyperfine splitting, the demands on the knowledge of the absolute beam energy is largely reduced to a few eV, with the exception of the lightest elements \cite{Geithner.2008,Nortershauser.2009,Krieger.2011}. However, potential changes during these relative measurements are critical and drifts of the acceleration voltage of $\approx 100$\,mV exceed the typical uncertainty, even for heavier isotopes.

Precision measurements of the rest-frame transition frequency $\nu_0$ can nevertheless be performed by probing the ion beam in collinear $(\uparrow \uparrow)$ \emph{and} anticollinear $(\uparrow \downarrow)$ geometry
\begin{equation}
    \nu_0 = \sqrt{\nu_{\mathrm{lab,}\uparrow \downarrow} \cdot \nu_{\mathrm{lab,}\uparrow \uparrow}}
    \label{eq:colacol}
\end{equation}
to become independent of the kinetic energy \cite{Riis.1986,Nortershauser.2009,Krieger.2012}. Again, drifts of the beam energy between the alternately performed collinear and anticollinear measurements need to be suppressed to avoid systematic deviations. Having extracted $\nu_0$ once, it can be used in turn to deduce the kinetic beam energy $E_\mathrm{kin}$ from a single measurement in collinear or anticollinear geometry
\begin{equation}
E_\textnormal{kin} =\frac{mc^2}{2}\frac{(\nu_0 - \nu_\textnormal{lab})^2}{\nu_0 \nu_\textnormal{lab}}~.
\label{eq:HV}
\end{equation}
This allows one to determine the offset between acceleration voltage and kinetic beam energy and reduces the corresponding uncertainty significantly \cite{Koenig.2021}. Such a beam-energy determination via the Doppler-shift method has also been applied previously to accurately measure high voltages \cite{Gotte.2004,Kramer.2018}.

Faster and technically easier than scanning the laser frequency $\nu_{\mathrm{lab}}$ is altering the ion velocity by applying a scan voltage to the laser-ion-interaction region. Since the laser frequency in the rest frame of an ion is then altered by changing the Doppler shift depending on the applied scan voltage, this procedure is called Doppler tuning. For a single resonance peak the typical scan range is $< 50$\,V and the systematic impact, e.g., due to field penetration, on the rest-frame frequency (Eq.\,\eqref{eq:colacol}) or the isotope shift is small as long as the measurements of the targeted isotopes and the reference isotope are performed at approximately the same scan voltage. This, however, requires the laser frequency to be changed between the measurements of different isotopes. 
Keeping the laser frequency constant for different isotopes or investigating isotopes with large hyperfine splitting requires to apply scan voltages between 100\,V and a few kV and field penetration into the interaction region becomes critical.

All measurements discussed in this work were performed at the collinear laser spectroscopy facilities at Michigan State University (MSU) and TU Darmstadt (TUDa). For a detailed description of the respective setups we refer to Refs.\cite{Minamisono.2013, Koenig.2020}. Here, we restrict the description to dedicated details of the recently installed high-voltage stabilization schemes.

\begin{figure*}[htb]
    \centering
    \includegraphics[width=1\linewidth]{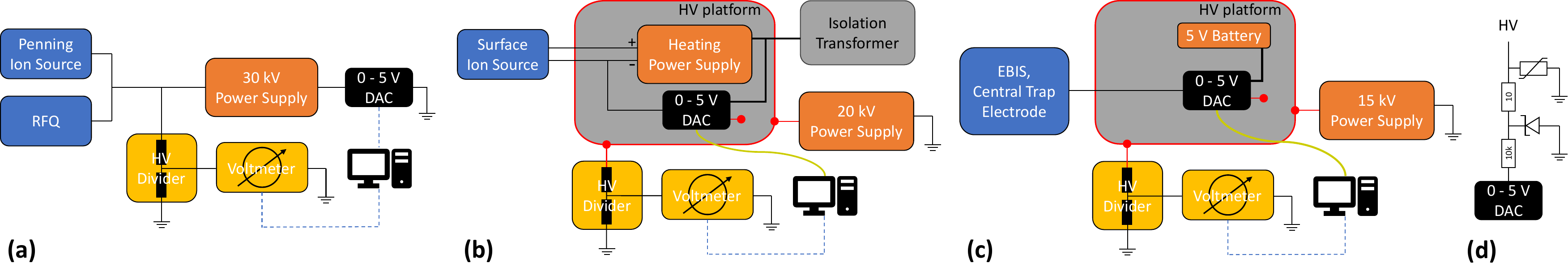}
    \caption{Voltage stabilization schemes at (a) MSU and (b,c) TUDa in combination with different ion sources. In all approaches, the applied high voltage is measured with a custom-made voltage divider and a digital voltmeter. A control loop is used to correct the high voltage with a 0--5\,V DAC unit. At MSU (a) the high-voltage power supply can be floated on this correction voltage, which is not possible at TUDa. Here (b,c), the DAC unit is placed on a high-voltage platform. Electrical connections to the platform are shown in red and communication with the DAC is realized via optical fiber (yellow). Using the surface ionization source (b) also requires a power supply that generates the heating current on the high-voltage platform, which is supplied with line voltage using an isolation transformer. The operation of the EBIS (c) does not require additional devices and the DAC can be supplied from a 5-V battery, which removes the voltage noise induced by the isolation transformer. To avoid malfunctions caused by high-voltage sparks, the DAC is protected with a varistor and a transient-voltage suppressor diode (d) in all cases.}
    \label{fig:Schaltung}
\end{figure*}

\subsection{MSU setup}
\label{sec:MSU-setup}
At the BEam COoler and LAser spectroscopy (BECOLA) facility at MSU, ion beams are available from the Facility for Rare Isotope Beams or from a local offline Penning ion source. The ions are first fed into a He-buffer-gas-filled radio-frequency quadrupole trap (RFQ) for beam cooling and bunching before collinear laser spectroscopy is performed. The offline source as well as the RFQ, including respective electronic devices, are floated with a FUG \textit{HCP 350 - 65 000 MOD} on 30\,kV. This voltage is measured with a high-voltage divider and a digital voltmeter (Keysight 34465A). The former commercial Ohmlabs HVS series divider was replaced by a new custom-made high-voltage divider described below, and the measured voltage was used as input parameter of a control loop to stabilize the total high voltage.
The FUG power supply enables the ground of the voltage generator to be separated from the common ground and, hence, allows us to apply a small correction voltage to float the FUG and to stabilize its total output voltage as shown in Fig.\,\ref{fig:Schaltung}a.
The small correction voltage is generated with a Labjack T4 digital-to-analog converter (DAC) while the FUG power supply was always kept at the same setpoint. As the drifts of the FUG did not exceed 2\,V, a control loop with the 0--5\,V DAC output of the Labjack was sufficient.

\subsection{TUDa setup}
At COALA (TU Darmstadt) different high-voltage power supplies are used for the employed surface ionization source (SIS) and electron beam ion source (EBIS). In both cases the voltage generator cannot be separated from the common ground of the device, inhibiting the direct floating by the correction voltage. Instead, the DAC unit to apply the correction voltage is placed on the high-voltage platform of the ion source.

In Fig.\,\ref{fig:Schaltung}b the setup of the SIS is shown, which is used to produce beams of elements with low ionization potentials, like alkaline and earth alkaline metals or lanthanides by directly heating a graphite crucible. The high voltage of up to 20\,kV is generated with a Heinzinger PNChp 20000-10pos that drifts with up to 40\,mV\,/\,20\,min, which corresponds to the time needed for a typical measurement. The voltage is applied to a high-voltage platform, housing the heating power supply and the DAC unit, and measured with a high-voltage divider and a Keysight 3458A multimeter. The ion-source voltage is corrected by connecting the DAC to the negative pole of the heating power. This eliminates the drifts and reduces the short-term voltage variations to 10\,mV in the range of seconds \cite{Mueller.2020}. The DAC is controlled via fiber from the measurement PC.

The EBIS at COALA is operated in continuous mode as this yields superior beam properties compared to pulsed operation \cite{Imgram.2023.PRA}. In this mode, the ions are produced in a trap with three electrodes at fixed potentials supplied by ISEG HM81001 power supplies that drift with up to 100\,mV\,/\,20\,min. For measurements in light, highly charged ions as, e.g., reported in Ref.\cite{Imgram.2023.PRL,Imgram.2023.PRA} for $^{12}$C$^{4+}$, the central electrode that mainly determines the beam energy is stabilized by placing the DAC unit on a high-voltage platform and feeding the correction voltage to the electrode, as depicted in Fig.\,\ref{fig:Schaltung}c. To avoid additional noise that is generally introduced when using an isolation transformer, the DAC unit was supplied by batteries. In the future, this system will be expanded to also control the end-cap electrode through which the beam leaves the source.

In the TUDa configuration with the DAC on the high-voltage platform, the commercial DAC unit did not sustain high-voltage sparks for long. 
A custom solution was designed, which is based on a 12-bit MCO4728 DAC with four channels. Matched to our capacity and high voltage, the first guard is an AMLV22 varistor combined with a 10\,$\Omega$ resistor. Residual currents are deflected through a SMBJ5.0A transient voltage suppressor (TVS) that shields the DAC together with a 10\,k$\Omega$ resistor as depicted in Fig.\,\ref{fig:Schaltung}d. Also, the power supply line of the DAC is connected to the identical guard setup. So far the DAC, supplied with such shielding, did not require any replacements. Furthermore, a serial-to-fiber connection was integrated on the custom-build board which allows a direct communication with the PC on ground potential.

\subsection{High-voltage divider setup}
\begin{figure}[h]
    \centering
    \includegraphics[width=0.4\textwidth]{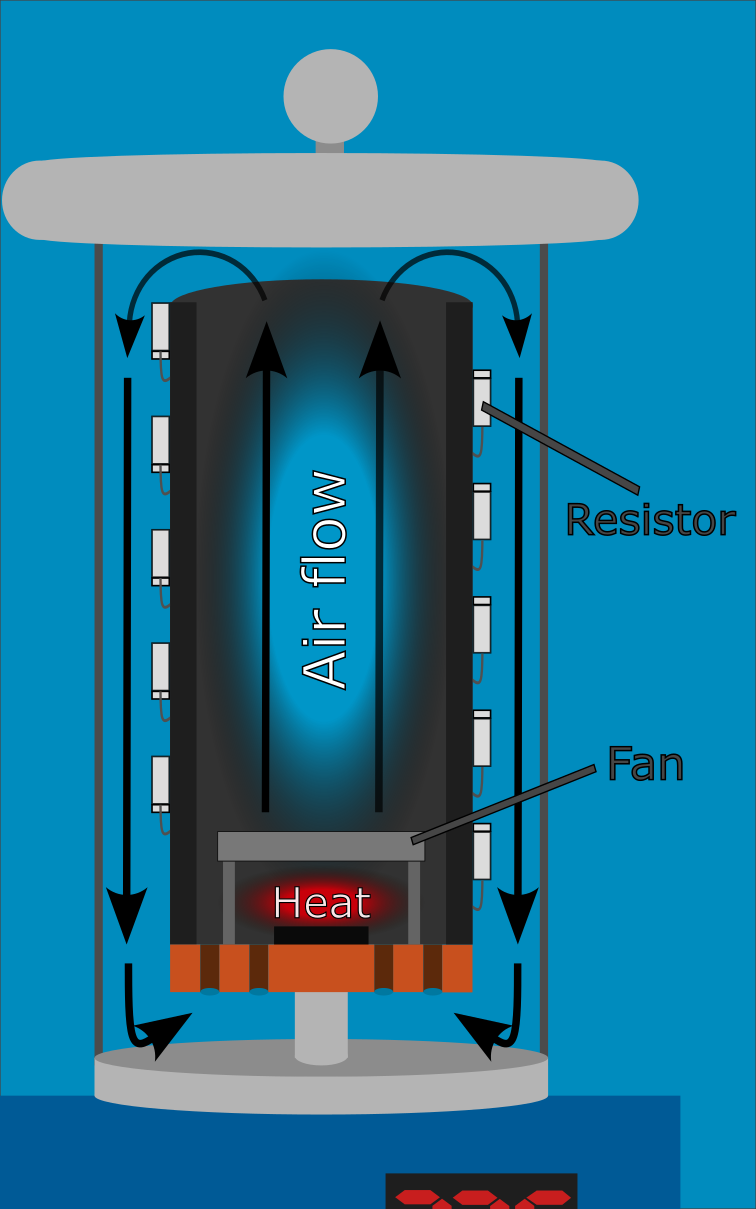}
    \caption{Cross section of the high voltage divider, illustrating the air flow in the temperature stabilized housing.}
    \label{fig:Heatflow}
\end{figure}

The high-voltage dividers used at MSU and TUDa are custom made and based on the identical design \cite{Passon.2024} and only differ in the divider ratio of 3000:1 and 2000:1, respectively, to scale the maximal applied voltage of 30\,kV and 20\,kV to 10\,V in both cases.
To match the requirement of a high temporal stability of the divider ratio, the dividers consist of a chain of resistors with a low-temperature coefficient of \SI[per-mode = symbol]{<1}{\ppm\per\kelvin}. This is achieved with Caddock film resistors of type USF370 with a resistance of \SI{9.95}{\mega\ohm} each that were individually characterized by the manufacturer and selected in sets of five to achieve a minimal total temperature coefficient. The maximal operation voltage is 1\,kV/resistor leading to a total resistance of 298.5 and \SI{199}{\mega\ohm}, respectively. As the low voltage resistor a Vishay foil resistor Y0007100K000T9L (\SI{100}{\kilo\ohm}, \SI[per-mode = symbol]{<2}{ppm \per K}) is used.
All resistors are placed inside a $\pm0.1$\,K temperature-stabilized housing to be independent from the ambient conditions. The resistor chain winds in a helix structure around a cylinder. This tower is surrounded by an acrylic glass tube and closed by metallic lids on the top and bottom. 
The basic principle of the temperature stabilization is shown in Fig.\,\ref{fig:Heatflow}. A thick copper plate in the center is heated to $30.0(1)^\circ$C and a ventilator creates a constant airflow that pushes up in the center and down on the outside, where the resistors are mounted. 
The heat created by the resistors is therefore equally distributed in the volume of the housing. 
As the temperature of the divider is set to a value above room temperature, there is passive cooling by the heat loss through the housing. 
The high voltage is applied to the top lid of the housing which has a toroidal shape to avoid discharges. The low-voltage end is at the bottom which is on ground potential. The electronics for the temperature stabilization are placed in a metallic box below, which also serves as the stand for the divider.

\section{Measurements}
\subsection{Calibration of the high-voltage divider}
\begin{figure}
    \centering
    \includegraphics[width=1\linewidth]{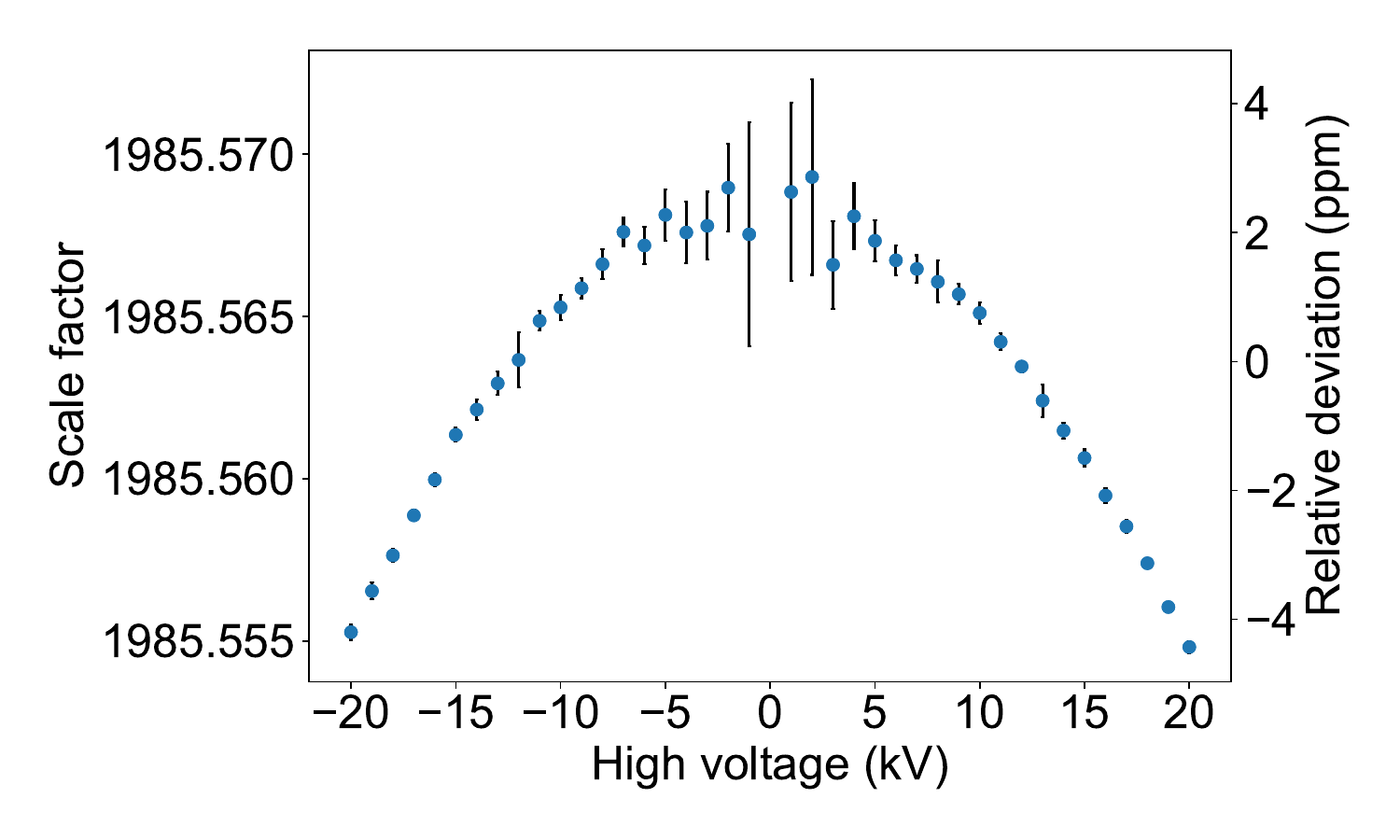}
    \caption{Scale factor of TUDa divider for different high voltages measured with the MT100 reference from PTB.}
    \label{fig:scale_factor}
\end{figure}

\begin{figure}
    \centering
    \includegraphics[width=1\linewidth]{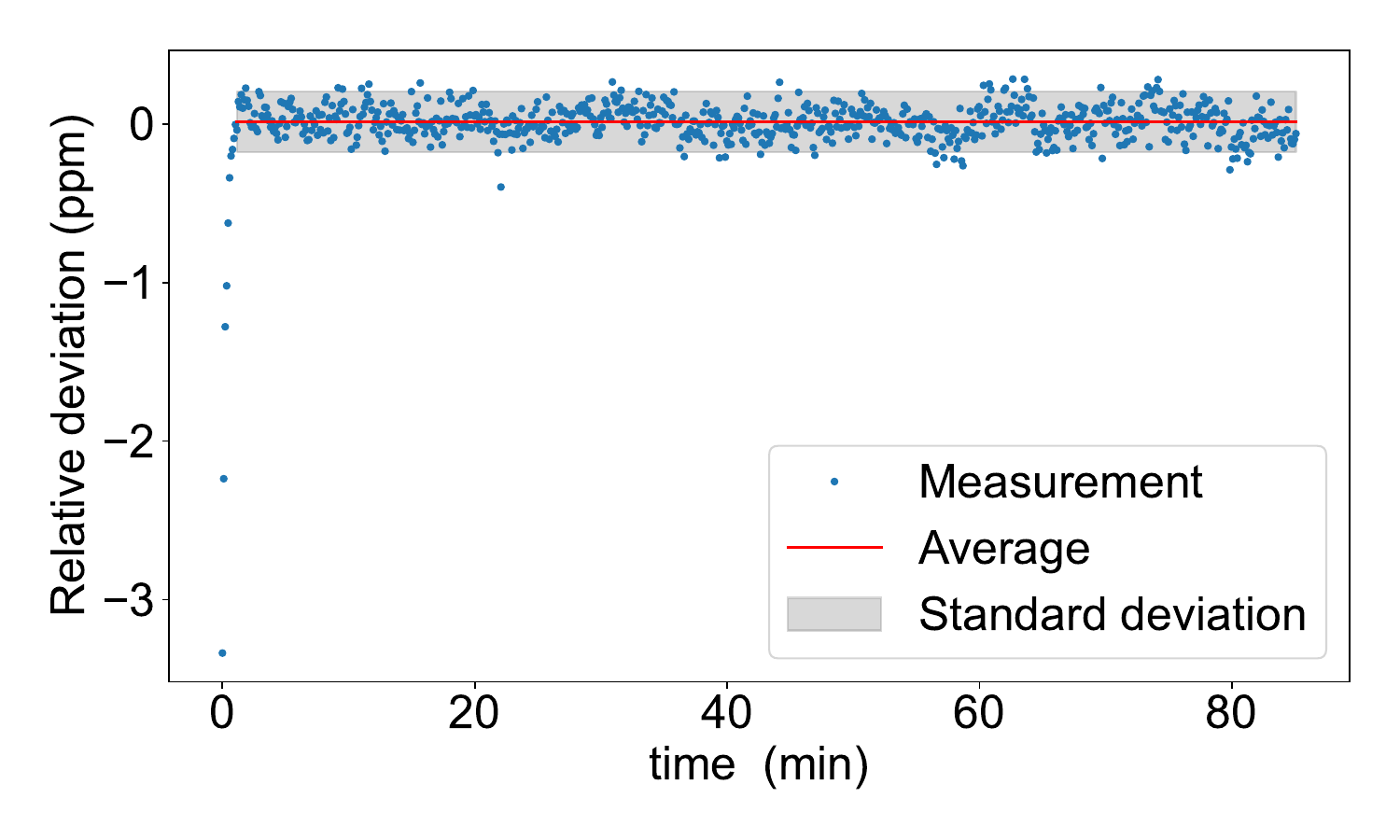}
    \caption{Relative deviation of the TUDa divider scale factor over time measured against the MT100 reference from PTB.}
    \label{fig:stability}
\end{figure}

To characterize a high-voltage divider and calibrate its scale factor, it has to be compared to a reference divider. At both facilities, this was done with a reference divider, MT100, from PTB \cite{Passon.2024}. 
Exemplary, we detail the calibration of the TUDa divider against the MT100 divider from PTB. The high voltage of interest is applied to both dividers and the voltage is recorded with two Keysight 3458A multimeters. This measurement is performed for several voltages and also for a short-circuited divider to determine contact voltages of the cables. The gain of the multimeters was calibrated with a Fluke 732C 10-V reference. Also here, the cable offset was considered.\\ 
As depicted in Fig.\,\ref{fig:scale_factor}, the scale factor of the TUDa divider was determined in steps of $\SI{1}{\kilo\volt}$ up to the maximal voltage of $\SI{20}{\kilo\volt}$. Repeating this process for both polarities leads to a parabolic shape of the data points. This is directly connected to the ohmic losses inside the divider. As the power of the losses $P_{\Omega}=U^2/R$ is directly proportional to the square of the voltage applied to the divider, the equilibrium temperature of the resistors is equally affected, which alters the resistance according to the temperature coefficient.\\ 
Besides the voltage dependence, also the long-term stability of the scale factor is important. To investigate this property, the maximum high voltage was applied to the divider and the PTB reference. In Fig.\,\ref{fig:stability} the temporal behavior of the relative deviation of the scale factor is shown. After about one minute the scale factor takes an equilibrium value that stayed constant over the whole measurement time of \SI{>80}{min}. The standard deviation indicated in Fig.\,\ref{fig:stability} is about $\SI{0.2}{\ppm}$ being well below the stabilization goal of \SI{<1}{\ppm}.  \\
The total uncertainty of the calibration of the scale factor at PTB is composed of the statistical uncertainty in Fig.\,\ref{fig:scale_factor} and the uncertainty of the reference divider MT100 of about $\SI{2}{\ppm}$ ($k=2$). Additionally, long-term drifts of the resistors need to be considered, which can change the scale factor by multiple \si{\ppm} pear year. Hence, this calibration process needs to be repeated regularly. 


\subsection{HV stability at FRIB}
\begin{figure}
    \centering
    \includegraphics[width=1\linewidth]{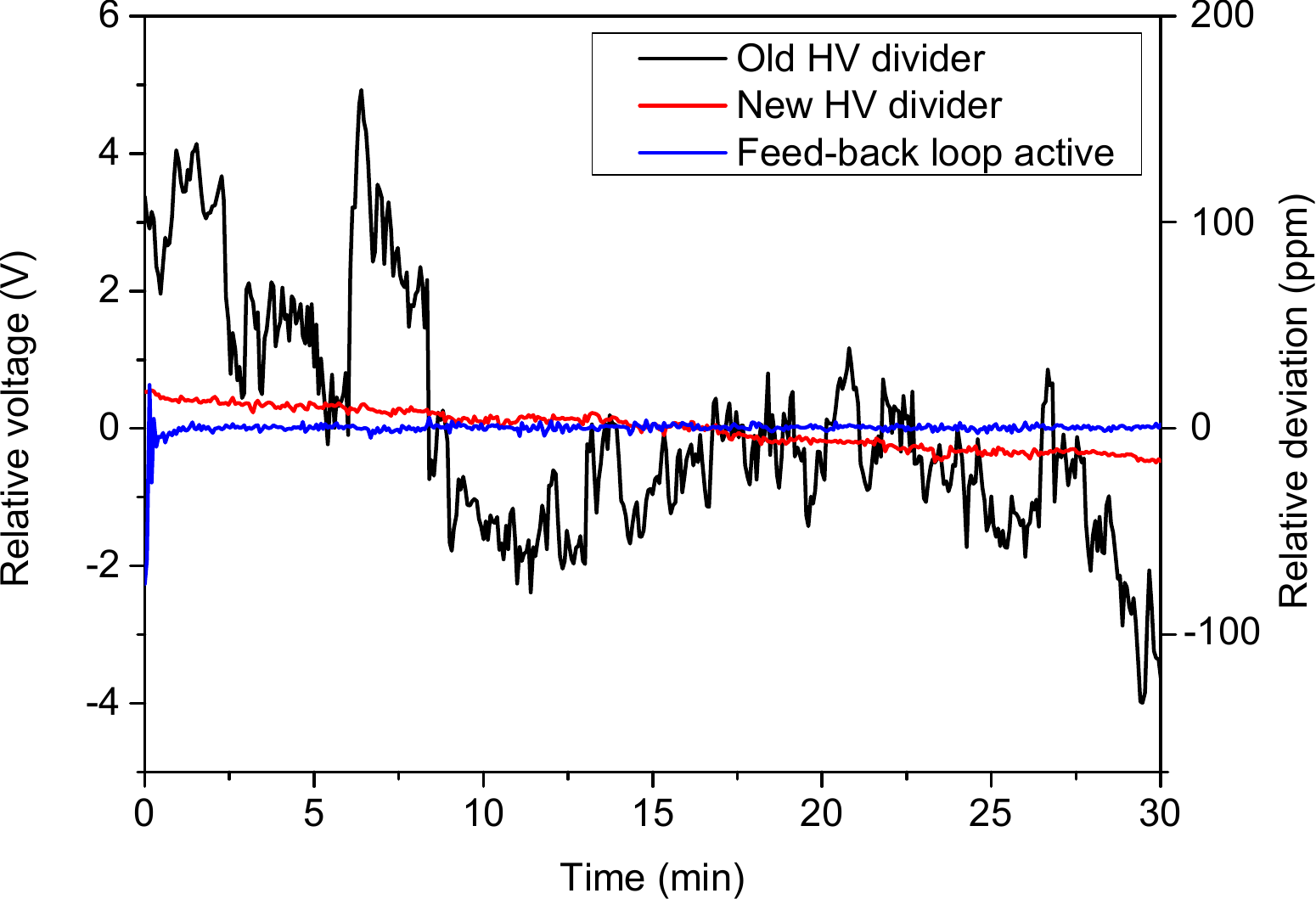}
    \caption{30-kV high-voltage measurement at BECOLA using a commercial high-voltage divider (black) and the present custom build divider (red) read out by 6.5 digits digital voltmeters. While the present divider showed a realistic slow drift of the high voltage, the commercial divider caused significant fluctuations of the voltage reading. Using the present divider and a feed-back loop, the slow, most-likely thermal drift of the power supply was compensated.    
    }
    \label{fig:FRIB-measurement}
\end{figure}

At FRIB, a high-voltage divider was built as described above to replace the former commercial Ohmlabs HVS series divider. Initially, the Ohmlabs divider performed well but then started to yield unrealistic readings. In Fig.\,\ref{fig:FRIB-measurement} the voltage applied with the ultra-stable FUG was measured with both dividers. While the new custom-built divider shows a small and steady drift of the high-voltage, which is commonly observed as power supplies tend to drift for hours after start to reach thermal equilibrium, the voltage measured with the old Ohmlabs divider shows unrealistic drifts and jumps. Using laser spectroscopy to quantify the voltage stability as described in the next section, also revealed that these large drifts are not real but rather an artifact of the divider's measurement. To eliminate the long-term drift of the high voltage, the stabilization scheme described in section\,\ref{sec:MSU-setup} was applied. As depicted in Fig.\,\ref{fig:FRIB-measurement}, the resulting voltage shows a high stability after an initial overshooting of the feed-back loop.

\subsection{HV stability TUDa}
\begin{figure}
    \centering
    \includegraphics[width=1\linewidth]{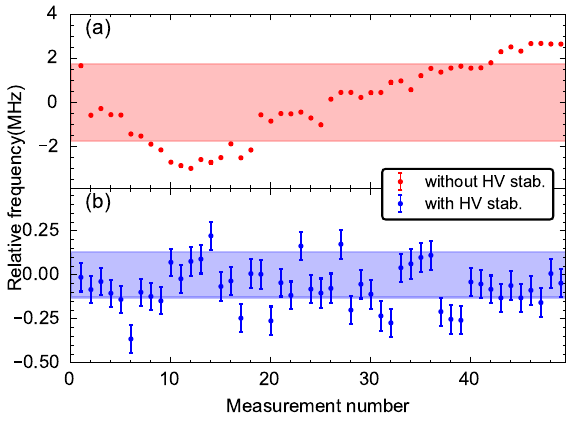}
    \caption{Stability of resonance frequencies of transitions in  $^{88}$Sr$^+$ with and without active high-voltage stabilization of the starting potential of the ions. Particularly, in the first hour after starting the power supply, the output voltage slowly settles, which was the case in (a). Due to the Doppler effect, the determined frequency changes if the starting potential and, hence, the ion velocity, is drifting. This can be circumvented when using the active high voltage stabilization as shown in (b).
    The shaded area corresponds to the standard deviation and decreases from 1.7\,MHz to 130\,kHz if the voltage-stabilization is active.}
    \label{fig:w/ and w/o}
\end{figure}

At TU Darmstadt a similar feed-back loop is employed to stabilize the high-voltage applied to the ion sources. Besides the electronic measurement, collinear laser spectroscopy of the  5s $^2$S$_{1/2}\rightarrow$ 5p $^2$P$_{J}$ transitions in $^{88}$Sr$^+$, produced in the surface ionization source, was employed to probe the high-voltage stability. For this purpose, 50 measurements at 20\,kV in anticollinear geometry were performed over a time period of 4\,h. Any voltage drift causes a change of the determined resonant laser frequency due to different Doppler shifts according to Eq.\,\eqref{eq:CLS}.
In Fig.\,\ref{fig:w/ and w/o}, the observed transition frequencies are plotted for the case without (top) and with (bottom) the stabilization scheme depicted in Fig.\,\ref{fig:Schaltung}b. In these measurements Doppler tuning was applied, as described above, and the laser frequency was locked to a reference cavity that was long-term stabilized to a frequency comb, achieving a frequency stability $< 100$\,kHz at the relevant time-scale, which is limited by the linewidth of the employed laser system \cite{Koenig.2020}.
As shown in Fig.\,\ref{fig:w/ and w/o}, the voltage stabilization significantly improved the reproducibility and reduced the standard deviation of the determined resonance frequencies from 1.7\,MHz to 130\,kHz. This is crucial to reach highest precision of rest-frame frequencies from collinear-anticollinear measurements and isotope shifts, e.g., to test atomic theory \cite{Mueller.2020} or to search for nonlinearities in a King-plot.


\section{Field penetration into the interaction region}
As discussed above, field penetration into the interaction region can have a substantial impact on the hyperfine-splitting and isotope-shift  measurements if Doppler tuning is applied. As their relation is in first order linear, it can be corrected in the data analysis if it is well quantified. Hence, the setups at MSU and TUDa were mapped accordingly.

\begin{figure}
    \centering
    \includegraphics[width=1\linewidth]{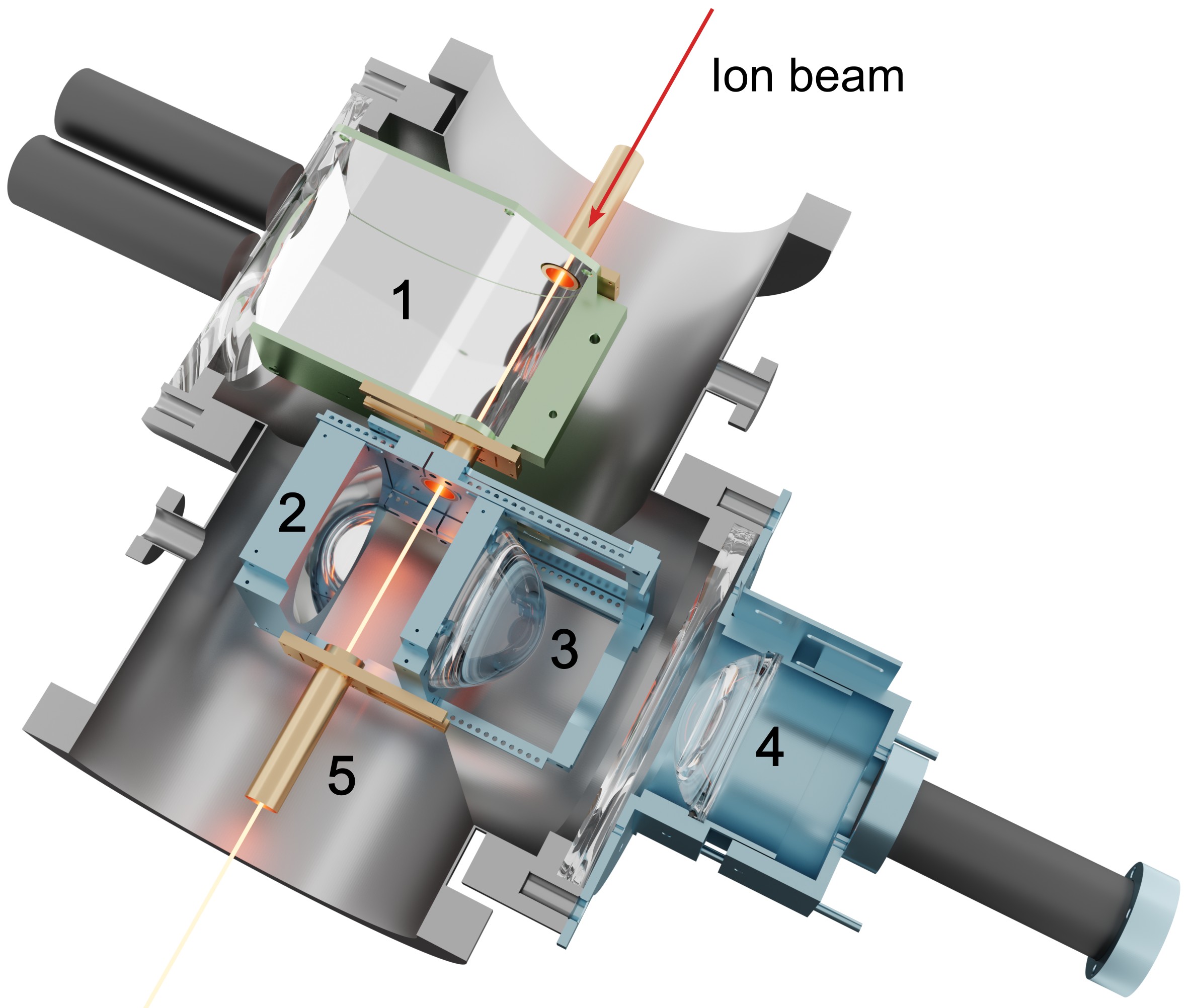}
    \caption{Fluorescence detection region of COALA at TUDa. In the first segment, an elliptical mirror (1) focuses fluorescence emitted from the beam axis onto two PMTs. In the second segment, a mirror (2) and two lenses (3, 4) focus fluorescence emitted from their focal point onto a single PMT. Small tubes (5) define the electrostatic potential outside and between the two segments and are used for the alignment of the FDR. }
    \label{fig:fdr}
\end{figure}

For the characterization of the FDR at TUDa, which is depicted in Fig.\,\ref{fig:fdr}, a single peak of the hyperfine spectrum of the $4f^{13}6s\rightarrow 4f^{13}6p$ transition at 384.0 nm in $^{169}$Tm$^+$ was used. First, collinear and anticollinear measurements were performed to extract its rest-frame frequency $\nu_0$ with Eq.\,\eqref{eq:colacol}. The laser frequencies were adjusted for the resonances to appear at the same scan voltage. Then, anticollinear measurements were performed with different laser frequencies so that the obtained resonance spectra spanned the complete scan-voltage range, and the kinetic beam energy was extracted with Eq.\,\eqref{eq:HV}. The corresponding laser frequency $\nu_\mathrm{lab}$ was measured with a frequency comb and the total uncertainty of the extracted effective voltage applied to the detection region was less than 0.05\,mV.
In parallel, the applied scan voltage was measured with a Keysight 3458A 8.5-digit multimeter.

\begin{figure*}
    \centering
    \includegraphics[width=0.9\linewidth]{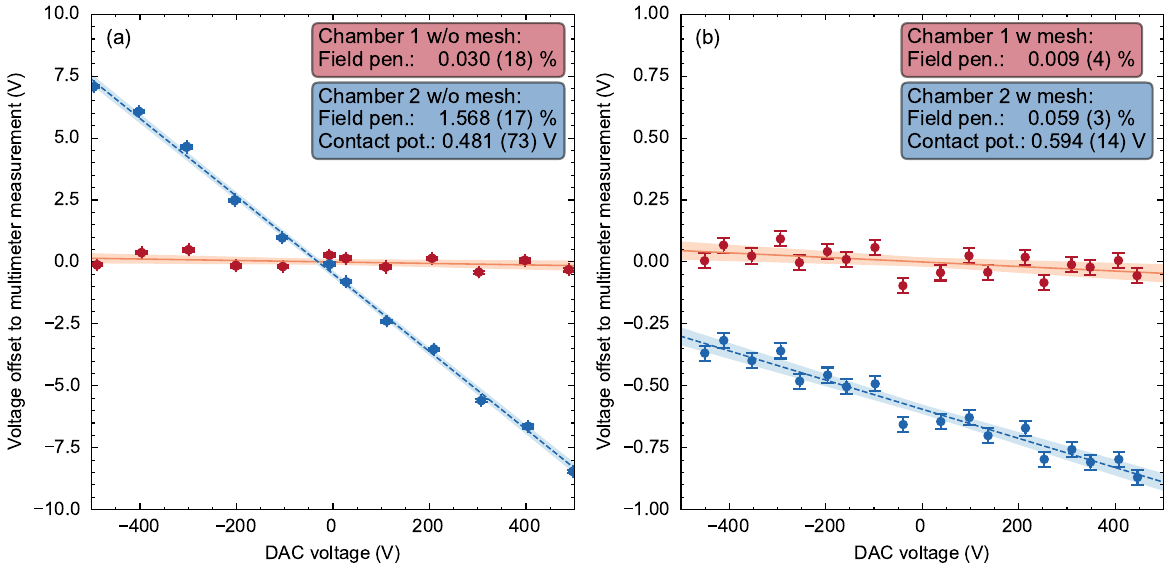}
    \caption{Difference of the laser spectroscopically determined potential applied to the detection region to an electronic measurement with a multimeter (a) without and (b) with additional shielding by installing an metallic mesh. Without mesh a large field penetration effect was observed particularly in chamber 2, which has a more open geometry than chamber 1 \cite{Mueller.2024}. Closing the openings with a metallic mesh reduced the field penetration. Between both chambers a potential offset is observed, which is ascribed to contact voltages as different materials are used.}
    \label{fig:Uscan}
\end{figure*}

In Fig.\,\ref{fig:Uscan}a the laser-based measurement of the total acceleration voltage of the ions is shown relative to the applied and electronically measured voltage for the two chambers of the fluorescence detection region. 
The first chamber is equipped with an elliptical mirror set that collects the fluorescence light at one focal point and guides it to the second focal point outside the vacuum chamber, at which the photomultiplier tube (PMT) is located \cite{Maass.2020}, see Fig.\,\ref{fig:fdr} (1). It has an enclosed geometry, i.e., it is only open towards the PMT and has a 13-mm diameter entry and exit port for the laser and ion beam. The entry and exit ports are connected with a 92-mm long and 13-mm wide tube.
Contrarily, chamber 2, which is equipped with a lens to focus the light onto the PMT \cite{Mueller.2024}, has a relatively open geometry without metallic shielding on the sides. This explains the significantly larger field penetration of 1.568\,(17)\,\% compared to 0.030\,(18)\,\% of chamber 1.
As different materials are used in the two chambers, different contact voltages exist causing an offset of almost 0.6\,V.

Afterward, the fluorescence detection region was improved by installing metallic meshes (3\,mm $\cdot$ 3\,mm openings, 1\,mm wire) at the opening toward the PMTs of chamber 1 and by enclosing the lens region of chamber 2 with the same mesh. The measurement was repeated, now using the 5s $^2$S$_{1/2}\rightarrow$ 5p $^2$P$_{1/2}$ transition in $^{88}$Sr$^+$. As shown in Fig.\,\ref{fig:Uscan}b, the field penetration was strongly reduced to 0.009\,(4)\,\% and 0.059\,(3)\,\% of chamber 1 and 2, respectively.
Deflections of the ion beam caused by the applied high voltage and misaligned electrodes would also lead to shifts of the resonant laser frequency (see Eq.\,\eqref{eq:CLS}). This would cause a parabolic function around 0 which was not observed. Similarly, a wrong gain factor of the multimeter would not cause a linear slope but an inversion at 0.

At MSU the same procedure was applied using the $3p^6 4s~^2\mathrm{S}_{1/2}\rightarrow 4p~^2\mathrm{P}_{3/2}$ transition (393\,nm) in $^{40}$Ca$^+$ but due to the absence of a frequency comb, a lower resolution was achieved. Here, a field penetration of 0.055\,(10)\,\% was observed in the first fluorescence detection chamber \cite{Minamisono.2013} while no field penetration could be resolved for the second chamber. The latter is a copy of the mirror-based chamber at TUDa with mesh. Hence, a penetration of 0.01\,\% is expected.
The relative field penetration between the two MSU chambers was confirmed in the measurement of hyperfine spectra in Sc$^+$ \cite{Dockery.2023}.
Generally, the quantification of the field penetration from the measurement of a well-known hyperfine splitting is the easiest approach since such a measurement can be performed at one laser frequency and no frequency comb is required to achieve a high resolution. Reference spectra can be obtained from well-characterized setups, e.g., at TUDa.

\section{Conclusion}
The determination of the kinetic beam energy, which is governed by the electrostatic acceleration potential, is crucial for precision fast-beam collinear laser spectroscopy measurements and constitutes the largest systematic uncertainty. With new custom-built high-voltage dividers the acceleration voltage at BECOLA and COALA can be precisely measured and stabilized to the ppm level with a feed-back loop. This eliminates the impact of voltage drifts during the experiment and enables the highest accuracy to be reached. Furthermore, the often neglected impact of field penetration into the optical detection region was investigated, yielding significant contributions if large Doppler tuning voltages are applied. This can be circumvented by performing all measurements at the same scanning voltage, which requires the change of the laser frequencies for different isotopes or hyperfine lines. Particularly for the latter, this can become impractical. With a quantification of the field penetration into the interaction region, either by performing measurements of the same transition at different scan voltages or easier by measuring a well-known hyperfine splitting, its impact can be attributed in the analysis. The relative uncertainty of such a calibration, here 100\,ppm and 40\,ppm for the setups at MSU and TUDa, respectively, is usually below the statistical uncertainty.

\section{Acknowledgments}
We thank Jörg Krämer, Phillip Imgram, and Hendrik Bodnar for their contributions to the experimental developments at COALA, and Uwe Bonnes for designing the custom DAC. We acknowledge funding by the Deutsche Forschungsgemeinschaft (DFG) Project No. 279384907 SFB 1245, as well as under Grant No. INST No. 163/392-1 FUGG, and NO 789/4-1. Furthermore, support was granted from the German Federal Ministry for Education and Research (BMBF) under Contract No. 05P21RDFN1 and 05P21RDCI1 and the National Science Foundation under Grant No. PHY-21-11185.


\section{Data availability}
The data that support the findings of this study are available from the corresponding author upon reasonable request.

\bibliographystyle{aipnum4-1}
\bibliography{literature.bib}

\end{document}